# MARTE Profile-based MDA approach for semantic NFP-aware Web services

**Hajar Omrana**
*Department of Computer Sciences, Ecole Mohammadia d'Ingénieurs (EMI),
Mohammad V University – Agdal
BP. 765 AV. Ibn Sina Agdal, Rabat, Morocco*
E-mail: hajaromrana@research.emi.ac.ma

**Fatima-Zahra Belouadha**
*Department of Computer Sciences, Ecole Mohammadia d'Ingénieurs (EMI),
Mohammad V University – Agdal
BP. 765 AV. Ibn Sina Agdal, Rabat, Morocco*
E-mail: belouadhaf@emi.ac.ma

**Ounsa Roudiès**
*Department of Computer Sciences, Ecole Mohammadia d'Ingénieurs (EMI),
Mohammad V University – Agdal
BP. 765 AV. Ibn Sina Agdal, Rabat, Morocco*
E-mail: roudieso@emi.ac.ma

### Abstract

Non-Functional Properties (NFPs) such as security, quality of service and business-related properties enhance the service description and provide necessary information about the fitness of its behaviour. These properties have become crucial criteria for efficient selection and composition of Web services. However, they belong to different domains, are complex, change frequently and have to be semantically described. The W3C standard WS-Policy, recommended to describe these properties doesn't define standardized specifications that cover all NFPs domains. Moreover, it doesn't provide an easy manner to express them independently of domains, and doesn't support their semantic. This paper proposes a Model-driven approach to describe and automatically generate enriched Web services including semantic NFPs. It explores both the use of the OMG Profile for Modelling and Analysis of Real-Time Embedded Systems (MARTE) and the W3C standards. Mapping rules, from NFPs profile to WS-Policy and SAWSDL files, transforms NFPs into policies associated with WSDL elements.

## 1. Introduction

A wide range of services are today available over the Web. Many of them offer similar functionalities and meet the same requirements. To select an appropriate service, the user should express his Non-Functional Properties (NFPs). While the description of functional properties (such as inputs, outputs, preconditions, effects, etc.) is mandatory to invoke a service, the specification of NFPs is necessary to select the appropriate service. In last few years, these properties have become crucial criteria to support efficient selection and composition of Web services, due to the huge number of available services offering similar functionalities. In this paper, we are interested to this aspect of Web services. The idea is to propose an MDA approach to model and automatically generate enriched Web services including semantic NFPs. The proposed approach should be aligned with the W3C standards and at the same time independent of the NFPs domains. In this context, we use and extend the MARTE (Modeling and Analysis of Real-Time Embedded Systems) NFPs profile to support modeling semantic NFPs. We propose both a meta-model of the W3C standard WS-Policy recommended to describe Web services policies and the WS-PolicyConstraints language used to specify the associated constraints, and we release the mapping from the MARTE NFPs profile to WS-policy and SAWSDL files to generate the enriched Web services.

The paper content is organized in five sections. Section 2 presents the problematic and the motivations of this work, especially, the limits of the WS-Policy standard and the NFPs' complexity. Section 3 describes our approach. It, firstly, provides an overview of the MARTE NFPs profile, which we explore and extend to remedy to the discussed problematic and support semantic annotations of NFPs. Then, it describes our WS-Policy meta-model and exposes a set of rules to perform models transformation from the MARTE NFPs profile to WS-Policy and SAWSDL files. Finally, it illustrates the use of this profile by applying it to a sample Web service in order to model its NFPs. Section 4 discusses the related works, and finally, Section 5 states the conclusions and future works.

## 2. Problematic and motivations

In the literature, there is no comprehensive list of NFPs, due to their dependency on user's domain and context. Nevertheless, several Web services approaches [1][2] have compiled lists of NFPs. The most notable work in this direction was presented by O'Sullivan et al. [2], which define a set of relevant NFPs (Service Provider, Temporal Model, Trust, Price, Security...) and describe their models using Object Role Modeling (ORM). However, several modeling approaches of NFPs in Web services context have met some limits due to the nature of these properties. Non-functional aspects are both abstract and delicate to formalize (the same property may have different implications in different contexts), numeric values can use different measurements units (e.g. 's', 'ms' and 'µs' as time units) and qualitative values depend on the user's domain (e.g. 'good', 'acceptable' or 'well'). As far as NFP constraints, they require a flexible language to support mathematical operators (e.g. '<', 'or', interval).

Besides, WS-Policy [3] is nowadays the most used standard for enriching WSDL files with NFPs, using a set of policies associated with the WSDL elements. However, several difficulties remain for defining the various assertions that are related to different domains (quality of service, security, business, etc.). On the one hand, existing WS-Policy specifications (such as WS-SecurityPolicy and WSReliableMessaging) do not cover all non-functional aspects (e.g. price, promotions, quality of service, etc.). On the other hand, WS-Policy does not specify a generic and domain-independent assertion language to represent NFPs and the associated constraints. Accordingly, providers and customers are taken to define their own WS-Policy specifications in order to represent NFPs related to their context and not covered by the WS-* family. Due to the perplexity and the abstraction of NFPs representation, it proves necessary to rely on domain experts rather than IT ones for modeling non-functional specific domains. However, domain experts don't have to deal with the WS-Policy grammar and would prefer using visual models.

As well, since WS-Policy specifications are not standardized, they have to be semantically annotated to allow matching offered and required policies when performing discovery, selection and contract negotiation agreement processes. Besides, the Web is a dynamic environment in which the services can change or become inactive. The Web services NFPs can, especially, be frequently updated by the providers. In this context, the main motivation of this work is to propose an approach compatible with the standards, adaptable to the changes and appropriate to specify and generate enriched Web services including semantic NFPs. The aim is to overcome the complexity of these properties and to provide a manner for describing them independently of domains.

## 3. Proposed approach

To remedy to the NFPs complexity and ensure a domain independent-modeling of these properties, we propose a MDA approach for designing Web services' semantic NFPs. Our approach uses the UML NFPs profile for Modeling and Analysis of Real-Time Embedded Systems (MARTE) [4] standardized by OMG, and its Value Specification Language (VSL) used to describe the NFPs constraints. At our knowledge, the use of MARTE NFPs profile in Web services context has never been investigated before. However, this remains very interesting.

Besides, to introduce NFPs semantic descriptions, we extend the MARTE NFPs profile using the SAWSDL tags. We also use a set of models transformation rules to automatically generate the corresponding WS-Policy and SAWSDL files. These proposed rules are used to map the NFPs profile's elements to WS-Policy standard and WS-PolicyConstraints ones [5].

In this section, we give an overview of the MARTE NFPs profile and present its extension to support semantic NFPs. Then, we describe our WS-Policy meta-model to which the modeled NFPs will be mapped. We finally expose a set of rules to perform models transformation from the MARTE NFPs profile to WS-Policy and SAWSDL files. We also present a case study "*FlightService*" using NFPs Profile to model the corresponding NFPs in order to validate and illustrate our approach.

### 3.1. MARTE NFPs Profile: Overview and semantic extension

MARTE NFPs sub-profile provides a modeling ground for semantically defining well-formed non-functional aspects. It allows describing basic, complex, quantitative or qualitative NFPs. In addition, it uses the VSL, a textual language, for expressing constraints and specifying NFPs values. The NFPs Profile is constituted of four main stereotypes: *Nfp*, *NfpType*, *NfpConstraint* and Unit. *Nfp* stereotype extends the UML Property. It specifies NFP information about one or more parameters articulating the behavior of the system such as quality of service, reliability, performance or trust [6]. *NfpType* stereotype defines the NFPs' data types. The *NfpConstraint* stereotype extends the UML Constraint. It represents a constraint concerning a NFP that must be satisfied by a real-time system. It is expressed using the VSL language. The Unit stereotype extends UML *EnumerationLiteral*. It associates a unit with a measured value of a NFP.

To semantically annotate the NFPs, we extend the MARTE NFPs profile using *SemanticNfp* stereotype. This stereotype extends *Nfp* stereotype of MARTE NFPs profile (Fig.1). Based on SAWSDL standard, it permits to attach one or multiple semantic concepts to a NFP using three attributes ModelReference, *LowringSchema* and *LiftingSchema* [7]. The ModelReference attribute provides information about the URI of a semantic concept that can be an ontological model element. The *LowringSchema* and *LiftingSchema* attributes provide information about the two-way mapping between an XML element describing an assertion and the corresponding semantic concept.

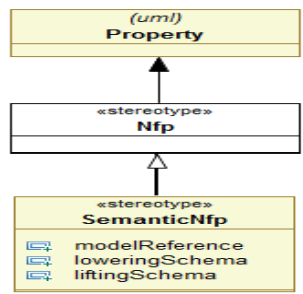

Figure 1: Extension of the NFP profile for describing Semantic NFPs

### 3.2. Our Web Service Policies Meta-Model

WS-Policy defines an abstract model for expressing the capabilities, requirements and general characteristics of the various entities in Web services policies form. Fig. 2 shows an enriched UML WS-Policy meta-model using the WS-PolicyConstraints specification and semantic elements.

According to WS-Policy standard, a policy may be related to a subject *PolicySubject* (an endpoint, message, operation, etc.) or a group of subjects *PolicyScope*. This standard defines operators to form policies based on the use of assertions and considers a policy as a collection of policy alternatives. Each policy alternative *PolicyAlternative* is a collection of assertions, and each assertion *PolicyAssertion* expresses a constraint, feature or requirement related to one or more vocabulary items. In our meta-model, we propose to model it as an aggregation of *PolicyConstraintsFunction*. According to WS-PolicyConstraints specification, which describes a generic and domain-independent language for expressing Web services policy assertion, each WS-PolicyConstraints function noted *PolicyConstraintsFunction* is a XACML predicate (Apply element) that represents a constraint. It can be applied over one or more vocabulary items. Each vocabulary item *VocabularyItem* represents a specific aspect that belongs to a policy domain *PolicyDomain*, has a data type and refers to a *SemanticElement*. This element would be defined by the domain expert.

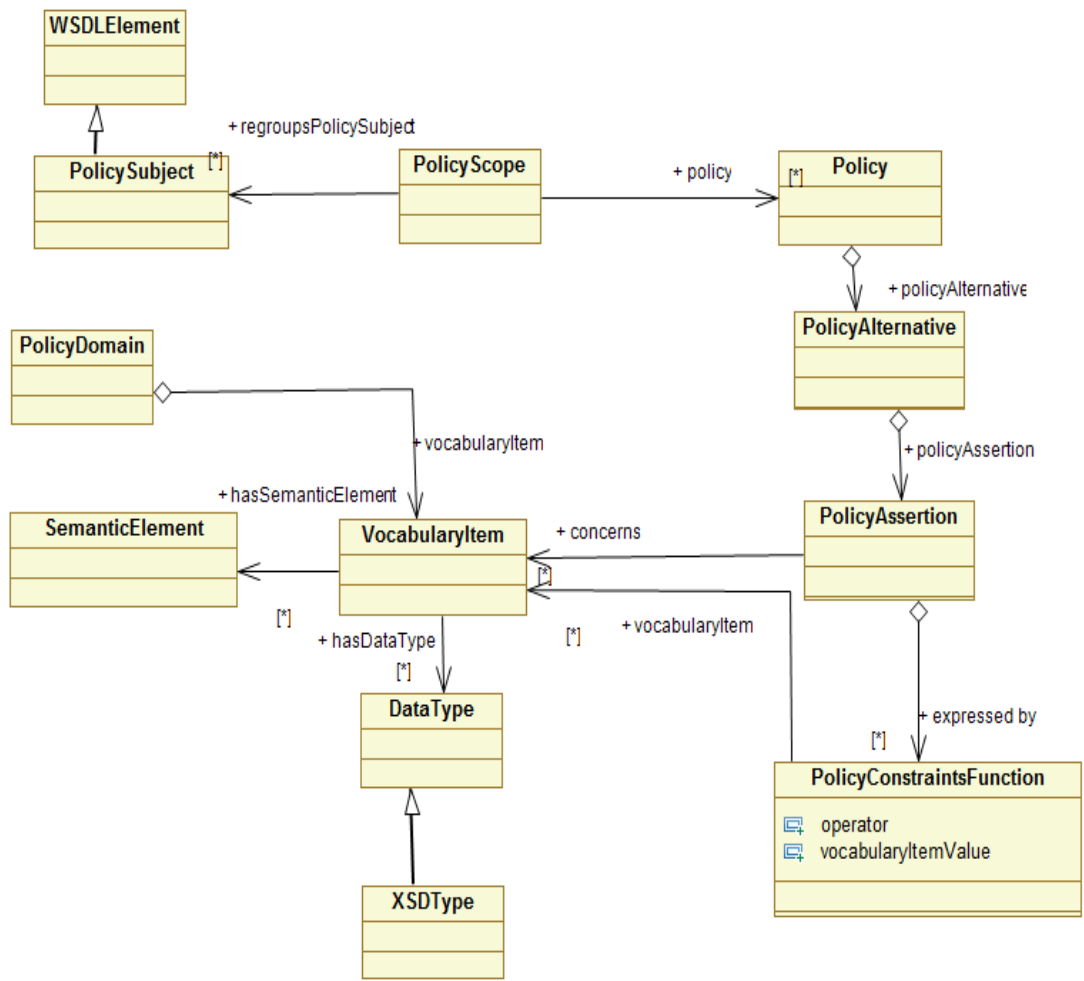

Figure 2: Web Services Policies Meta-Model

### 3.3. Transforming NFPs Profile elements to Web Services Policies files

This section describes the transformations we propose to generate executable web services' specifications enriched by NFPs from MARTE NFPs UML profile. They concern transformations from NFPs Profile' elements to WS-Policy Meta-Model (Model to Model transformations) and transformations from WS-Policy Model to XSD, WS-Policy, WS-PolicyConstraints and SAWSDL files (Model to Text Transformations). As a first attempt, we focus on simple VSL constraints and basic NFPs data types and operators.

Table 1 shows examples of mapping between the elements of the MARTE NFPs profile and the classes of Web service policy meta-model. We consider a NFP as a vocabulary item specific to particular non-functional aspect. Thus, the stereotype *Nfp* is mapped to *VocabularyItem* class and *NfpType* stereotype is mapped to *XSDType* class. Each basic NFPs type from MARTE Library is mapped to an appropriate Xml Schema Description (XSD) type. *NfpConstraint* is translated to a policy containing alternatives policies and assertions which correspond to the concerned constraint. The constraint's operator is expressed by WS-PolicyConstraints function operator and the constraint's literal value by WS-PolicyConstraints function *VocabularyItem*. The offered *ConstraintKind* is mapped to a policy which contains an empty alternative policy tag (</wsp: All >).

**Table 1:** Examples of NFPs Profile / WS-Policy Meta-Model Mapping

| MARTE NFPs Profile' Element | Type | WS-Policy Meta-class |
|---|---|---|
| Nfp | Stereotype | VocabularyItem |
| SemanticNfp | Stereotype | SemanticElement |
| NfpType | Stereotype | XSDType |
| NfpConstraint | Stereotype | Policy/ PolicyAlternative/ PolicyAssertion |
| Or Aggregation | VSL Expression | Policy |
| And Aggregation | VSL Expression | PolicyAlternative |
| Constraint's Operator | VSL Expression | PolicyConstraintFunction .operator |
| Constraint's literal value | VSL Expression | PolicyConstraintFuction.vocabularyItem |
| ConstraintKind Offred | VSL Expression | Empty PolicyAlternative |

Mapping step aims to generate SAWSDL and WS-Policy files with embedded WS-PolicyConstraints functions from WS-Policy meta-model. Fig. 3 shows an example of transformation rules described in an informal way only to show the coherences between our meta-model and WS-PolicyConstraints function.

```
<Apply FunctionId="wspc&function,
                  [PolicyConstraintsFunction.operator]">
<AttributeValue DataType="&xsd;
                PolicyConstraintsFunction.VocabularyItemId.XSDType]">
             "[PolicyConstraintsFunction.vocabularyItemValue]"
</AttributeValue> <ResourceAttributeDesignator
AttributeId="[PolicyConstraintsFunction.VocabularyItemId.name]"
DataType="&xsd;[PolicyConstraintsFunction.VocabularyItemId.XSDType]"/>
</Apply>
```

**Figure 3: Sample Transformation Rule for WS-PolicyConstraints Function**

### 3.4. Case Study

To illustrate our approach and the use of the MARTE NFPs profile, we consider a *FlightService* offered by a travel agency. This service provides useful information about a given flight. Our main objective is to represent NFPs associated with *FlightService*'s WSDL elements: Service and Endpoint, and generate the appropriate files using the developed rules.

Fig. 4 shows both an extract of the class model related to *FlightService* using the proposed WSDL 2.0 profile described in our previous work [8], and the corresponding generated SAWSDL and WS-Policy files. Two NFPs are assigned to the service element stereotyped webService: Price whose data type is NFP_Price and Availability whose data type is *NFP_Percentage*. Besides, the NFP Delay is applied to the Endpoint element and has as data type *NFP_Real*. All defined data types (*NFP_Price*, *NFP_Percentage* and *NFP_Real*) are extended data types defined in the *Basic_NFP_Type* sub-package of the MARTE Library.

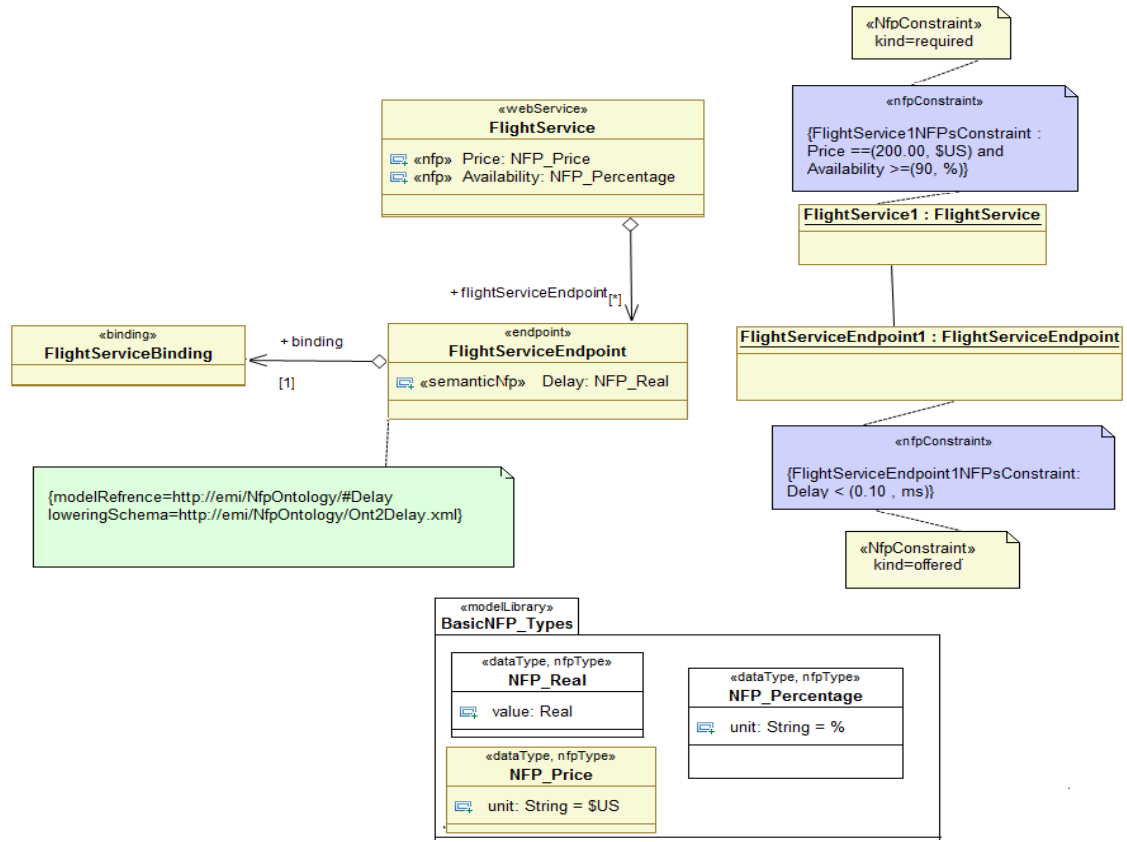

**Figure 4: FlightService's NFPs Model using the MARTE NFPs Profile**

Since NFPs are often dynamic and vary over time, their values and associated constraints will be generated whenever a *FlightService*'s WSDL document is requested. The corresponding object diagram displays the constraints and NFPs values at the time of the WSDL request. FlightService1 and FlightServiceEndpoint1 instances are respectively bound to the required FlightService1NFPsConstraint and the offered FlightServiceEndpoint1NFPsConstraint. They indicate the values, units and associated constraints corresponding to the NFP Price, NFP Availability and NFP Delay attributes. FlightService1NFPsConstraint requires a price equal to 200.00 $US and an availability greater than 90 %. Otherwise, FlightServiceEndpoint1NFPsConstraint requires a delay lower than 0.10 ms. When Price and Availability (*FlightService* class's attributes) are stereotyped *Nfp*, Delay (*FlightServiceEndpoint* class's attribute) is stereotyped *SemanticNfp* to semantically annotate it by referring to an ontological concept. Each NFP and the associated VSL constraints has been transformed to Web service policies attached to Service and Endpoint elements of *FlightService*'s WSDL. The generated XSD file is imported into <wsdl: types> as shown is Fig 5.

We note that the main emphasized properties of our work are:
- we investigate the use of NFPs sub-profile from the MARTE UML Profile for modeling Web services NFPs and introduce WS-PolicyConstraints to make the description of NFPs constraints domain-independent,
- we propose an approach compatible with the W3C standards WS-Policy and SAWSDL used to describe Web services semantic NFPs,
- we address the necessary transformations to generate executable WS-Policy, XSD and SAWSDL files, and finally,
- we implement a practical example to assess the feasibility of our approach.

```xml
<wsdl:service name="FlightService1"
               Interface="FlightServiceInterface"

<wsp:Policy   wsu:Id= "FlightService1NFPsPolicy" >
    <wsp:All>
        <wsp:Policy>
         <wsp:All>
         <Apply FunctionId="wspc&function;double-equals">
         <AttributeValue DataType="&xsd;double">200.00</AttributeValue>
         <ResourceAttributeDesignator
                 AttributeId="Price/PriceValue" DataType="&xsd;double"/>
         </Apply>
         <Apply FunctionId="wspc&function;string-equals">
             <AttributeValue DataType="&xsd;string"> $US </AttributeValue>
               <ResourceAttributeDesignator
                AttributeId="Price/PriceUnit" DataType="&xsd;string"/>
         </Apply>
        </wsp:All>
        </wsp:Policy>
        <wsp:Policy>
         <wsp:All>
            <Apply FunctionId="wspc&function;integer-greater-than-or-equals">
           <AttributeValue DataType="&xsd;integer"> 90 </AttributeValue>
               <ResourceAttributeDesignator
         AttributeId="Availability/AvailabilityValue" DataType="&xsd;integer"/>
           </Apply>
           <Apply FunctionId="wspc&function;string-equals">
           <AttributeValue DataType="&xsd;string"> % </AttributeValue>
              <ResourceAttributeDesignator
            AttributeId="Availability/AvailabilityUnit" DataType="&xsd;string"/>
           </Apply>
        </wsp:All>
        </wsp:Policy>
<  <wsp:All>
</wsp:Policy>
 <wsdl:endpoint name="FlightServiceEndpoint1"
                binding="FlightServiceBinding1"
                adresss="http://emi.ac.ma/projects/WebServices">
 <wsp:Policy wsu:Id= "FlightServiceEndpoint1NFPsPolicy" >
 <wsp:ExactlyOne>
        <wsp:All>
           <Apply FunctionId="wspc&function;double-less-than">
           <AttributeValue DataType="&xsd;double">0.10</AttributeValue>
           <ResourceAttributeDesignator
                  AttributeId="Delay/DelayValue" DataType="&xsd;double"/>
           </Apply>
           <Apply FunctionId="wspc&function;string-equals">
           <AttributeValue DataType="&xsd;string"> ms </AttributeValue>
           <ResourceAttributeDesignator
                    AttributeId="Delay/DelayUnit" DataType="&xsd;string"/>
           </Apply>
        </wsp:All>
         <wsp:All />
 </wsp:ExactlyOne>
 </wsp:Policy>
 </wsdl:endpoint>
</wsdl:service>
```

Figure 5: Extract of SAWSDL file content of FlightService

## 4. Related Works

Nowadays, many research works are interesting in Web services technologies [15, 16]. They are focusing on modeling, discovery, selection, static or dynamic composition and semantics of Web services. In the Web services modeling field, to which our work relates, the MDA approach has been exploited in several works. Taking into consideration the recommendations of different standards such as WSDL, OWL-S and SAWSDL, the aim of this category of works was to automatically generate interface files describing, especially, functional, technical and semantic aspects of Web services.

Nevertheless, Web services' NFPS should also be regarded as an essential aspect that must be considered in all Web services' related activities such as discovery, selection or dynamic composition. They often constitute technical requirements, quality attributes or simply system' properties covering all aspects that cannot be embedded in functional parameters. In this section, we are particularly interested in works that cover Web services non-functional aspects using WS-Policy. We especially discuss those that proposed UML profiles and transformation models to express Web services' policies.

In the literature, we found two works [9] [10] that can be considered as close related works. Ortiz and Hernandez [10] propose a UML profile based on the aspect oriented approach. The idea is to attribute dynamic and flexible behaviors to Web services NFPs and Policy implementations based on aspect-oriented techniques. This profile allows attaching alternatives and constraints (extra NFPs) to the interface of a Web service and its operations. However, it does not allow the developer to describe the non-functional aspects of endpoints and bindings. In comparison with our work, they don't define any model or language to design the extra NFPs. This makes modeling specific NFPs, related to a given context, a complex operation for the user. Jegadeesan et al. [9] apply a model driven systematic approach to develop Web service policies. The current approach addresses different levels of policy aspects (technical, service-level and domain-level aspects). It proposes the description of service policies using visual models. It also discusses the transformation of these policy models to executable files using standards as WS-Policy and WS-PolicyConstraints. This work doesn't cover the relationship between non-functional aspect and policies and doesn't support semantic descriptions.

Concerning non-functional vocabulary specifications, several Web services approaches [2][11][12] have addressed specific lists of NFPs for a given service. We think that the most relevant among them is this that was proposed by O'Sullivan et al. [2]. It includes various types of NFPs such as Service Provider, Temporal Model, Trust, Price and Security. To model these categories, authors use Object Role Modeling (ORM) and also produce concrete a XML syntax which can be considered as a vocabulary reference for NFPs. The NFPs taxonomies and vocabulary elaborated in the other works [11][12] are, however, less detailed.

Besides, other authors present a Web service's selection approach based on NFPs [13]. This approach uses WS-Policy standard and the Ontology Web Language (OWL) to describe semantic NFPs of Web services. However, this approach does not consider the dynamic aspect of NFPs that change over the time. A meta-model for describing NFPs for Web services has also been proposed in another work [14]. The approach distinguishes the NFP offered by vendors from those requested by users. It uses a Policy-based model in order to specify the NFPs. Nevertheless, the expressiveness of the associated constraints remains limited.

## 5. Conclusion and outlook

In this paper, we have investigated the use on MARTE NFPs profile to model Web services' NFPs. This has the following main advantages: First, our solution can be considered as an independent and open modeling framework for basic and complex NFPs. Second, it increases the level of abstraction through which the domain's expert can define its own non-functional vocabulary domain according to a given context. Finally, it uses a flexible language (VSL) to define constraints over these properties and an interesting library of measurements units, NFPs types, etc. However, the MARTE profile does not support semantic annotations of NFPs. To make it possible to semantically describe these properties and then favorite automatic selection and composition processes, we extended it using the SAWSDL elements.

To make our approach compatible with the W3C Web services standards, we developed transformations rules in order to map NFPs profile into a WS-Policy meta-model. In this meta-model, we considered a set of elements according to WS-Policy standard and also explored the flexibilities offered by the WS-PolicyConstraints language. The use of this language allows expressing NFPs constraints using domain-independent assertion language. In this context, we also proposed a mapping from VSL expressions to WS-PolicyConstraints functions..

Besides, to evaluate our approach, we used a sample practical example FilightService. The generated file validates the proposed meta-model and transformations rules. This lets conclude that the MARTE NFPs profile and the VSL language are also adapted to model Web services NFPs and the associated constraints independently of domains. Another conclusion of this work consists in the possibility to semantically annotate the Web services NFPs using SAWSDL Language. This standard is essentially recommended to annotate functional elements and XSD. However, since the non-functional assertions are modeled as XML schema in our approach, it was be possible to annotate them using the same standard (SAWSDL). The considered annotations can be embedded in WSDL files in a manner fully compatible with the W3C standards.

Finally, we think that the proposed approach would minimize the cost of developing and updating NFPs in a dynamic environment as the Web, and overcome the complexity of the services standard specifications. Our future works would be in the following directions: (a) developing transformation rules to map complex VSL expressions to WS-PolicyConstraints and covering all NFPs types offered by MARTE (b) developing a semantic matching algorithm to enable the semantic-based research [17] and (c) exploring MARTE Generic Quantitative Analysis Modeling (GQAM) and Performance Analysis Modeling (PAM) UML sub-profiles in Web services context.